\documentclass[superscriptaddress,citeautoscript,aps,prl,preprint,footinbib]{revtex4-1}

\usepackage{graphicx}
\usepackage{enumitem}
\usepackage{gensymb}
\usepackage{amsmath}
\usepackage{amssymb}
\usepackage{dcolumn}
\usepackage{color}
\begin{document}

\title{Small polarons and the Janus nature of $\text{TiO}_\text{2}$(110)}

\author{Ji Chen}
\email{ji.chen@pku.edu.cn}
\affiliation{School of Physics, Peking University, Beijing 100871, P. R. China}
\affiliation{Department of Physics and Astronomy, London Centre for Nanotechnology, Thomas Young Centre, University College London, Gower Street, London WC1E 6BT, U.K.}
\affiliation{Max Planck Institute for Solid State Research, Heisenbergstrasse 1, 70569 Stuttgart, Germany}

\author{Christopher Penschke}
\affiliation{Department of Physics and Astronomy, London Centre for Nanotechnology, Thomas Young Centre, University College London, Gower Street, London WC1E 6BT, U.K.}

\author{Ali Alavi}
\affiliation{Max Planck Institute for Solid State Research, Heisenbergstrasse 1, 70569 Stuttgart, Germany}
\affiliation{Department of Chemistry, University of Cambridge, Lensfield Road, Cambridge CB2 1EW, United Kingdom}

\author{Angelos Michaelides}
\email{angelos.michaelides@ucl.ac.uk}
\affiliation{Department of Physics and Astronomy, London Centre for Nanotechnology, Thomas Young Centre, University College London, Gower Street, London WC1E 6BT, U.K.}
\affiliation{Max Planck Institute for Solid State Research, Heisenbergstrasse 1, 70569 Stuttgart, Germany}

\begin{abstract}
Polarons are ubiquitous in many semiconductors and have been linked with conductivity and optical response of materials for photovoltaics and heterogeneous catalysis,
yet how surface polarons influence adsorption remains unclear.
Here, by modelling the surface of rutile titania using density functional theory, we reveal the effect of small surface polarons on water adsorption, dissociation, and hydrogen bonding.
On the one hand the presence of such polarons significantly 
suppresses dissociation of water molecules that are bonded directly to polaronic sites.
On the other hand, polarons facilitate water dissociation at certain non-polaronic sites.
Furthermore, polarons strengthen hydrogen bonds, which in turn affects water dissociation in hydrogen bonded overlayer structures. 
This study reveals that polarons at the rutile surface have complex, multi-faceted, effects on water adsorption, dissociation and hydrogen bonding, highlighting the importance of polarons on water structure and dynamics on such surfaces.
We expect that many of the physical properties of surface polarons identified here will apply more
generally to surfaces and interfaces that can host small polarons, beyond titania.
\end{abstract}

\maketitle

The fundamental physics of defects at clean surfaces and aqueous interfaces 
is of great importance to everyday processes such as wetting and nucleation, as well as the more application oriented topics of photovoltaics and heterogeneous catalysis \cite{fujishima_tio2_2008, berger_electrochemistry_2012, kuhlenbeck_well-ordered_2013, sosso_crystal_2016}.
Atomic defects on metal oxides, such as oxygen vacancies and interstitials have received most interest, and are often considered to increase surface activity \cite{onda_wet_2005,thompson_surface_2006, di_valentin_electronic_2006, Wendt_role_2008, henderson_surface_2011, katz_electron_2012, pang_structure_2013, paier_oxygen_2013, wang_localized_2015, wen_defects_2018,yin_excess_2018}.
For example, on the surfaces of $\text{TiO}_\text{2}$ -- a key material for photocatalysis
and photovoltaics and also a widely studied model system \cite{fujishima_tio2_2008, zhou_novel_2016, pham_modelling_2017, wen_electronic_2018, atambo_electronic_2019} --
surface oxygen vacancies are active sites for water dissociation 
\cite{diebold_perspective:_2017},
with implications for water splitting as a potential means to produce abundant and clean hydrogen fuel \cite{campbell_introduction:_2013}.

Polarons are another important class of defects that 
have recently been highlighted and linked to the activity of titania \cite{selcuk_facet-dependent_2016, yim_visualization_2018, reticcioli_interplay_2019, gaberle_role_2019}, ceria \cite{paier_oxygen_2013} and haematite \cite{katz_electron_2012, carneiro_excitation-wavelength-dependent_2017}.
Polarons are quasi-particles consisting of an electron bound to a lattice distortion, and in a small polaron this coupling is localized to within one or two lattice spacings.
Polarons can be considered as the charge carrier and mid-gap trapping state in the perspective of electronic band theory \cite{Shluger_1993,spreafico_nature_2014}.
They are particularly important in metal oxides, where they can be generated through photoabsorption,
or via reducing defects such as oxygen vacancies and cation interstitials \cite{thompson_surface_2006, Wendt_role_2008, henderson_surface_2011, katz_electron_2012, pang_structure_2013, paier_oxygen_2013, carneiro_excitation-wavelength-dependent_2017}.
Recently, several studies on $\text{TiO}_\text{2}$ have shown that 
the surface segregation and dynamics of polarons can be observed in real space  \cite{aschauer_influence_2010, kowalski_charge_2010, setvin_direct_2014, yim_engineering_2016, selcuk_facet-dependent_2016, yim_visualization_2018, reticcioli_formation_2018},
and that surface polarons and adsorbates influence each other.
For example, on the (110) surface of rutile $\text{TiO}_\text{2}$, where polarons appear as small polarons, 
Yim et al. found that water induced the segregation of polarons to the rutile (110) surface
\cite{yim_visualization_2018}. 
In addition, on anatase $\text{TiO}_\text{2}$, first principles calculations have predicted 
that excess electrons in the substrate tend to localize towards surface hydroxyl groups formed upon water dissociation \cite{selcuk_facet-dependent_2016}.

The above observations raise the question of how water is affected by small polarons on the rutile (110) surface.
This is an important question because even on well-prepared samples, the structure and dissociation states at the aqueous 
interface of rutile (110) is still under debate, see e.g.  Refs. \cite{YATES20091605,serrano_2015,hussain_structure_2017,diebold_perspective:_2017, balajka786}.
Water dissociation is often affected by specific atomic defects such as oxygen vacancies as mentioned above, impeding conclusive answers for the intrinsic property of the stoichiometric $\text{TiO}_\text{2}$~\cite{schaub_oxygen_2001, brookes_imaging_2001, wendt_formation_2006, bikondoa_direct_2006, zhang_water_2009, kristoffersen_role_2013}.
For example, photoelectron spectroscopy and diffraction studies show evidence for a dissociation channel on the five-fold-coordinated Ti ($\text{Ti}_\text{5c}$) site \cite{walle_experimental_2009, duncan_water_2012}, suggesting water dissociates without defects.
But measurements by Wang et al. and other microscopic surface experiments are in favor of water not dissociating on non-defective rutile (110) surfaces \cite{wang_probing_2017, diebold_perspective:_2017}.
Considering the fact that polarons are major charge carriers in such systems \cite{kowalski_charge_2010, setvin_direct_2014,yim_engineering_2016, yim_visualization_2018}, establishing the general role of polarons to water could bridge the gap between our knowledge on the pristine surfaces and surfaces with various specific defects.

In this study, we employ density functional theory (DFT) to investigate the role of polarons at the water-rutile (110) interface.
We find significant site dependence for both intact and dissociated water adsorption, 
leading to
direct suppression of water dissociation on polaronic sites and an
indirect enhancement of dissociation at non-polaronic adsorption sites.
We also examine how polarons influence the hydrogen bonds (HBs) between adsorbed water molecules and discuss
how this indirectly affects water dissociation.
Overall, we see that polarons have a surprisingly rich influence on the properties of wet titania. 
This behaviour is likely to hold for other adsorption systems and interfaces, and indeed we show that methanol on $\text{TiO}_\text{2}$ (110) is also strongly influenced by polarons.
The polaronic effects identified in this study could be relevant to interfacial structure, photocatalytic efficiency of many materials beyond titania.

DFT calculations were carried out with the VASP \cite{kresse_efficient_1996} and CP2K/QUICKSTEP \cite{vandevondele_quickstep:_2005} \text{ab initio} simulation packages.
The primary means by which polarons were introduced into $\text{TiO}_\text{2}$ was by adding an excess electron (with a compensating background charge).
However, calculations in which polarons were introduced by removing an oxygen atom were also performed.
With such a model we considered a polaronic configuration where water molecule adsorbs on the surface polaronic site instead of the vacancy site.
To ensure that a well-defined polaron structure forms, we occasionally 
elongated specific Ti-O bonds (by 0.03 \AA) prior to our geometry optimizations.
Tuning the initial structure in this manner directs the excess electron towards a specific Ti atom, and facilitates polaron formation after full geometry optimization.
Full details of the computational set-up employed are given in the supporting information (SI) \cite{si}, which includes Refs. \cite{kresse_efficient_1996, vandevondele_quickstep:_2005, kresse_ultrasoft_1999, heyd_hybrid_2003, heyd_erratum:_2006, perdew_generalized_1996, dudarev_electron-energy-loss_1998, goedecker_separable_1996, VandeVondele_Gaussian_2007, guidon_auxiliary_2010, PhysRevLett.110.095505, liu_structure_2010,cococcioni_linear_2005,kowalski_charge_2010,setvin_direct_2014,selcuk_facet-dependent_2016, grimme_consistent_2010, grimme_dispersion_2011, klimes_chemical_2010, klimes_van_2011, pang_structure_2013, matthiesen_formation_2009,yim_visualization_2018, lee_water_2013}. 
Since previous computational studies have shown that the properties of titania surfaces can be highly sensitive to the computational settings (see e.g. Refs. \onlinecite{pacchioni_modeling_2008, cramer_density_2009, angelis_theoretical_2014}),
care was taken to ensure that the key conclusions reached here were not dependent 
on the details of set-up used (see the SI) \cite{si}.
%

\begin{figure}[htb]
\begin{center}
\includegraphics[width=3.0in]{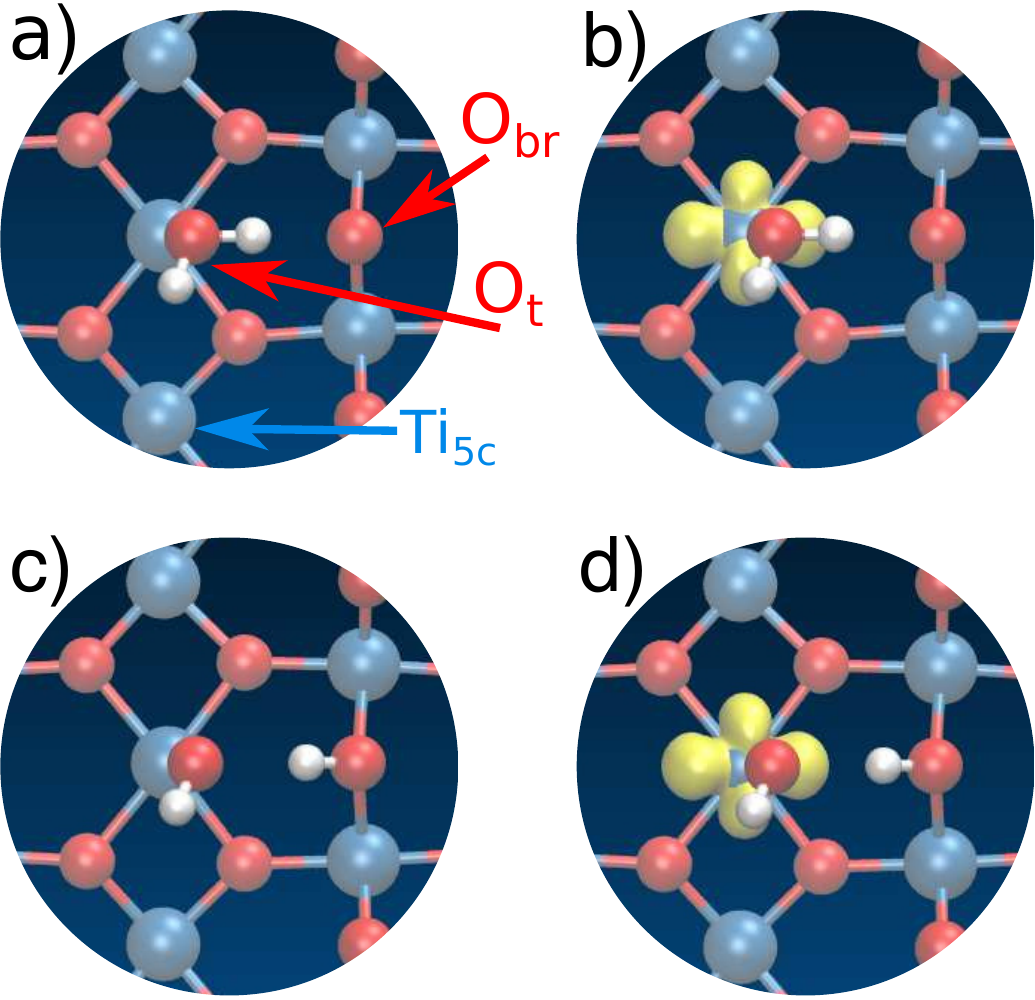}
\end{center}
\caption{
Water monomer adsorption structures on rutile (110). (a) and (b) are adsorbed structures of a molecular water on the pristine and the polaronic substrate, respectively.
(c) and (d) show the dissociated water species.
The yellow isosurfaces show the charge density of the polarons.
Note that for clarity only a small portion of the unit cell used in the calculations is shown.
}
\label{figure1}
\end{figure}

\begin{figure}[htb]
\begin{center}
\includegraphics[width=3.5in]{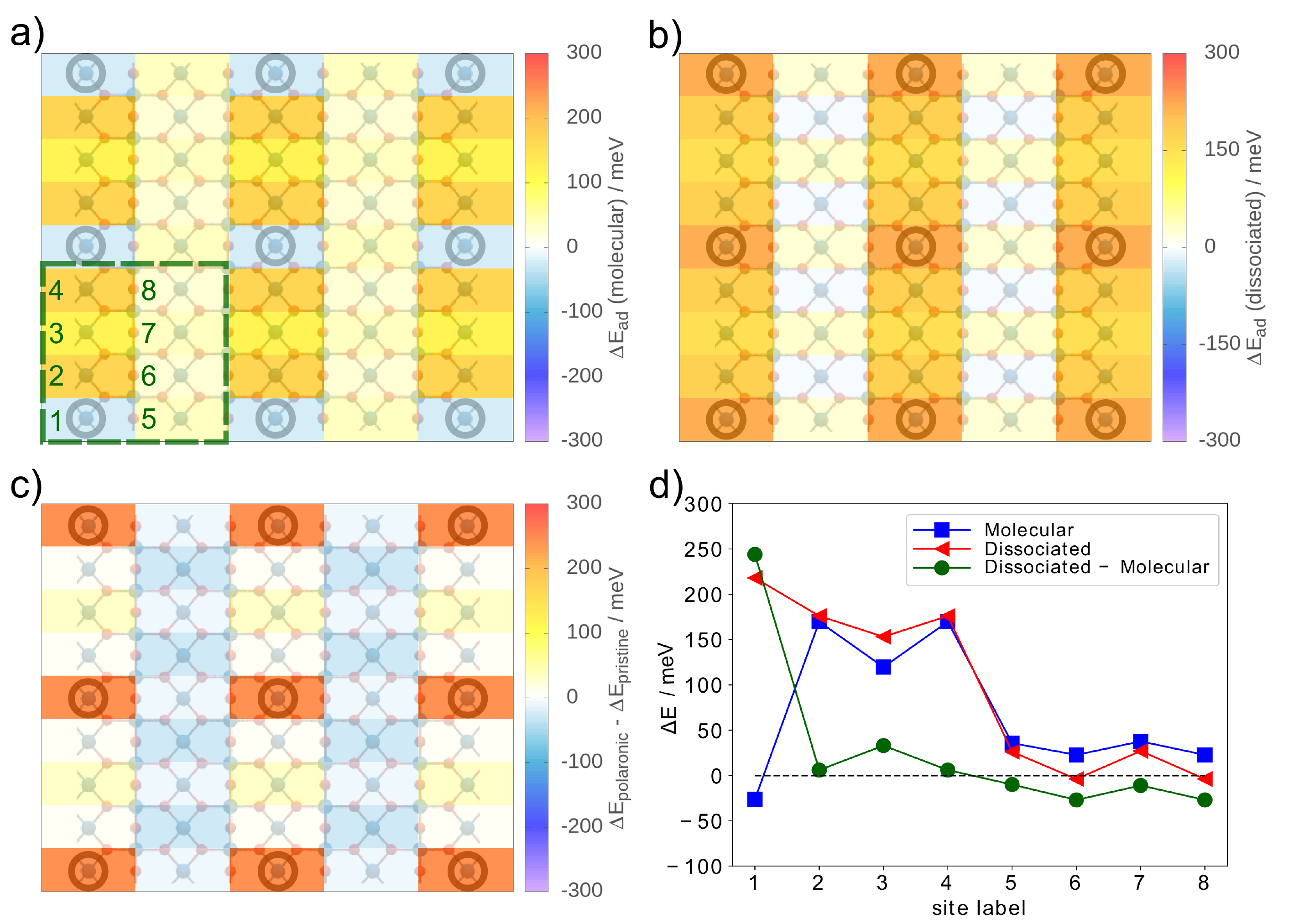}
\end{center}
\caption{The presence of localized polarons breaks the equivalence of the adsorption sites on the  rutile (110) surface.
(a) and (b) Heat maps of the change in adsorption energies for molecular and dissociated water, respectively, upon introducing a polaron.
Positive values indicate that polarons suppress adsorption.
(c) Heat map showing the effect polarons have on the water dissociation energy.
Positive values indicate that dissociation is suppressed. 
In all three heat maps the polarons are indicated with the black
circles and periodic $2 \times 4$ unit cells are indicated 
with the green dashed box in panel (a). 
(d) Plots showing the effect polarons have on the adsorption energy of molecular water and dissociated water, and the dissociation energy on different sites as indicated in panel (a).
}
\label{figure2}
\end{figure}

We begin by looking at the adsorption of an individual water molecule 
on $\text{TiO}_\text{2}$ (110).
The most stable adsorption configuration is shown in Fig. \ref{figure1}a, where the water 
molecule binds on top of the five-fold coordinated titanium ($\text{Ti}_\text{5c}$) and forms a HB with a bridging oxygen ($\text{O}_\text{br}$). 
This adsorption structure agrees with many previous theoretical and experimental studies (see e.g. Refs. \onlinecite{henderson_surface_2011, pang_structure_2013, diebold_perspective:_2017} and references therein).
The adsorption energy of the intact water molecule on the pristine surface, with the specific computational set-up employed, is ca. -0.92 eV.
Fig. 1b shows a water molecule adsorbed directly on a polaronic $\text{Ti}_\text{5c}$ site.
The yellow lobe corresponds to the charge density of the gap state, showing
that the polaron is a small polaron localized on the $\text{Ti}_\text{5c}$-$3d$ orbital (also see Fig. S2).
On the polaronic site the adsorption energy of the water molecule is slightly increased compared to the pristine surface to ca. -0.94 eV.
The presence of a localized polaron breaks the equivalence of the adsorption sites and in the neighborhood of the polaron a range of water adsorption energies is found. 
This is shown by the heat map of water molecule adsorption energies on  $\text{Ti}_\text{5c}$ sites in Fig. \ref{figure2}a.
Interestingly, it can be seen from  Fig. \ref{figure2}a that water adsorption on non-polaronic sites can be significantly destabilized, with the largest suppression up to ca. 170 meV.
Overall we see that water molecule adsorption at the polaronic site is favoured over the non-polaronic sites.

Dissociation of a water molecule leads to the adsorption configuration of Fig. \ref{figure1}c, consisting of
a bridging hydroxyl group ($\text{O}_\text{br}\text{H}$) and a terminal hydroxide group ($\text{O}_\text{t}\text{H}$).
On the same polaronic substrate as discussed above, the $\text{O}_\text{t}\text{H}$ can adsorb on the polaronic site (Fig. 1d) or on non-polaronic sites.
The site-dependent change in the $\text{O}_\text{t}\text{H}$ adsorption energy is shown in Fig. \ref{figure2}b.
Compared with the adsorption of an intact water molecule, the adsorption of dissociated water is affected quite differently by the polaron. Specifically, (i) adsorption on the polaronic site is significantly suppressed; and (ii) adsorption on non-polaronic sites is also suppressed but to a smaller extent.
This means that on a polaronic substrate, a dissociated water molecule shows an energetic preference for non-polaronic sites over polaronic sites.

Considering the effects of polarons on molecular water and dissociated water together, we obtain a site-dependent dissociation energy, defined as the energy difference between the dissociated state and
the molecular state (Fig. \ref{figure2}c).
On the polaronic sites, water dissociation is very strongly suppressed by ca. 240 meV. 
However, we also find that water dissociation at non-polaronic sites 
can be affected either negatively or positively, by up to ca. 30 meV.
Therefore, we see that polarons present two faces to the dissociation of water. 
On the one hand, a water molecule adsorbs preferentially on the polaronic site and dissociation is inhibited.
On the other hand, there are non-polaronic sites on a polaronic substrate where water dissociation is enhanced. 
%

\begin{figure}[htb]
\begin{center}
\includegraphics[width=3.5in]{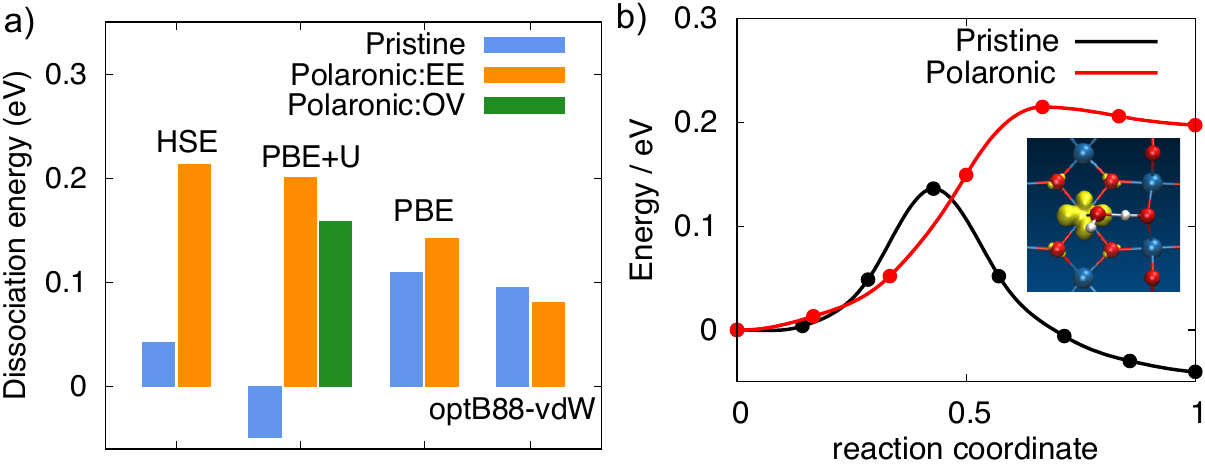}
\end{center}
\caption{
(a) The dissociation energy for a water monomer on $\text{TiO}_\text{2}$ using different exchange correlation functionals and models.
The dissociation energy is defined as the energy difference between the dissociated and molecular adsorption states. EE and OV indicate that the polarons are induced by an excess electron and an oxygen vacancy, respectively.
(b) The energy profiles for water dissociation calculated with PBE+U 
on pristine and polaronic $\text{TiO}_\text{2}$ surfaces, computed with the nudged elastic band approach.
}
\label{figure3}
\end{figure}

The suppression of water dissociation on the polaronic site is a direct effect of electrostatic interaction because both the polaron and the $\text{O}_\text{t}\text{H}$ are negatively charged. 
Therefore, correctly accounting for the localization of electrons is essential to establish the role of polarons.
The results above are based on a particular DFT flavor, namely the Perdew-Burke-Ernzerhof plus the Hubbard correction (PBE+U) \cite{dudarev_electron-energy-loss_1998};
results with U = 4.2 eV are reported, for results with other U values see the SI.
In Fig. \ref{figure3}a we show calculations with other DFT functionals.
First we consider the Heyd-Scuseria-Ernzerhof (HSE) hybrid exchange correlation functional \cite{heyd_hybrid_2003, heyd_erratum:_2006}, which can also describe the local nature of the polaron.
As with PBE+U, HSE also shows a significant suppression of water dissociation at the polaronic site.
The PBE and optB88-vdW (a van der Waals inclusive functional \cite{klimes_chemical_2010, klimes_van_2011}) calculations in Fig. \ref{figure3} are considered as 
control sets.
These functionals fail to describe the localized nature of the polaron and
as a result, the 
suppression of water dissociation is not correctly predicted.
In addition, the large suppression role on polaronic sites holds for polarons created by means of
surface oxygen vacancies.
Although water dissociates at surface oxygen vacancies, when the polaronic $\text{Ti}_\text{5c}$ sites 
adsorbed away from the vacancy dissociation is suppressed (Fig. \ref{figure3}).
This shows the suppression role is a physical effect of polarons and does not depend on how polarons are created.
%

Furthermore, we find the suppression of water dissociation by polarons not only raises the dissociation energy,
but also increases the dissociation barrier.
Fig. \ref{figure3}b plots the energy profile along the dissociation path of water on a pristine substrate and on a polaronic substrate calculated using the climbing image nudged elastic band method (cNEB) \cite{henkelman_climbing_2000}.
On the pristine substrate the energy barrier is 0.14 eV, while in the presence of polarons the barrier increases
to 0.22 eV.
There is also a significant reduction of the recombination barrier, from 0.18 eV to ca. 0.01 eV.
Thus there is an almost negligible barrier to reverse dissociation on a polaronic surface site.

More often than not, water on surfaces adsorbs at higher coverages than isolated monomers, forming a rich variety of hydrogen bonded structures \cite{sosso_crystal_2016}. 
To understand the influence of polarons in this regime 
we considered a range of models from low to high water coverage, including monomers, dimers, a monolayer, and a liquid water film (detailed in the SI). 
Fig. 4a summarizes the dissociation energy on the pristine and polaronic substrates for all models studied. 
First, we see that there is a suppressing effect of polarons for all 
adsorption structures considered. 
Second, when the coverage is low and adsorbed water molecules are not hydrogen bonded to each other, polaronic effects 
are rather similar to the monomer scenario discussed above, featuring a large suppression of water dissociation.
However when HBs form between water molecules, the influence of polarons is more subtle and intimately connected with the nature of the HB. 

\begin{figure}[htb]
\begin{center}
\includegraphics[width=3.3in]{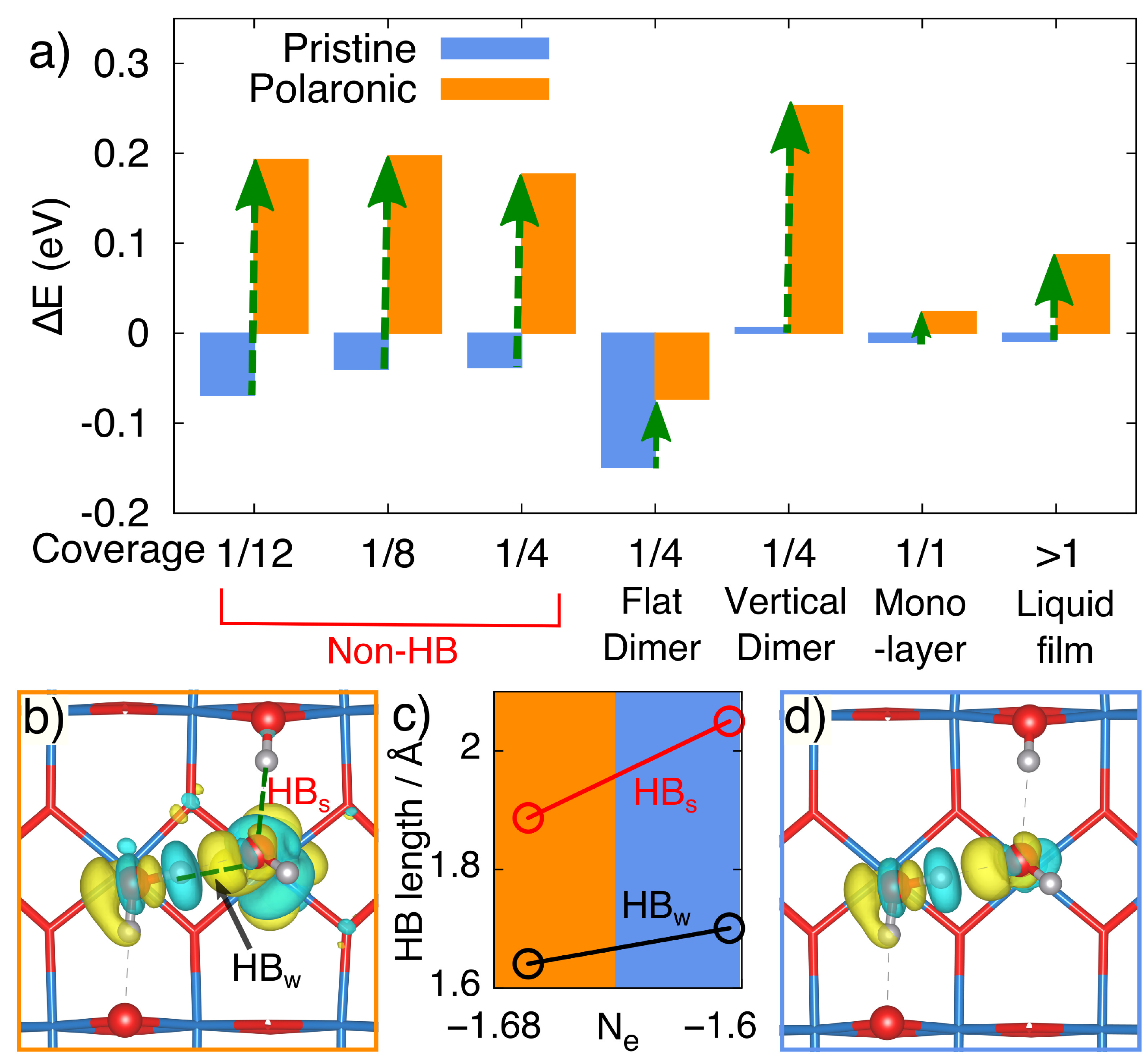}
\end{center}
\caption{
(a) The effect of polarons on water dissociation on surfaces with different water coverage.
(c) The length of HBs and the valence charge on the pristine surface (blue) and the polaronic surface (orange).
The HBs are shown with the green dashed lines in panel (b).
(b) and (d) are the charge rearrangement upon HB formation shown with the isosurfaces ($\pm$0.002 $e \cdot Bohr^{-3}$) of $\Delta\rho$ 
on the polaronic surface and the pristine surface, respectively.
$\Delta\rho=\rho_{surface+dimer}+\rho_{surface}-\rho_{surface+water_1}-\rho_{surface+water_2}$ \cite{michaelides_simulating_2007}.
$\rho_{surface+water_1}$ and $\rho_{surface+water_2}$ are charge densities of the surface with the first and the second water molecule, respectively. 
}
\label{figure4}
\end{figure}

Here we discuss the effect of polarons on hydrogen bonding, which not only explains the complex behavior identified at high water coverage but also
on its own is essential to e.g. structure, stability and dynamics of water on surfaces.
Using the flat dimer structure as an example (Fig. \ref{figure4}b,d), 
dissociation leads to the formation of an $\text{H}_2\text{O}$-OH group.
In this adsorption complex there are two HBs (one internal and one between the OH and the bridging hydroxyl).
Both are affected by the polaron.
Specifically, as shown in Fig. \ref{figure4}c, they are shortened and hence strengthened by polaron doping.
Analysis shows that this is because the valence charge on the terminal oxygen increases, making it a better
HB acceptor.
Fig. \ref{figure4}b,d shows the charge rearrangement upon the formation of the HB within $\text{H}_2\text{O}$-OH group on the polaronic and the pristine surface, respectively \cite{michaelides_simulating_2007}.
The slight enhancement in the charge rearrangement also indicates the strengthening of the HB.
For the molecular water dimer, HBs are enhanced by polarons as well, but to a much lesser extent. 
Therefore, the fact that polarons enhance HBs in $\text{H}_2\text{O}$-OH more than HBs in the water dimer effectively diminishes the suppression effects to water dissociation.
%

The observation that polarons increase HB strengths also applies to other models studied here.
For example, the HB between the water monomer and the surface bridging oxygen shrinks from 1.76 \AA~ to 1.71 \AA~ for molecular water and from 2.02 \AA~to 1.62 \AA~ for the dissociated structure upon addition of polarons.
For more complicated systems, such as the vertical dimer (a dimer structure extracted from the liquid),
the monolayer, and the liquid film models, the suppression effects on water dissociation and the enhancing effects on HBs play off against each other.
The overall effects are sensitive to the specific models employed, but qualitatively for all models studied 
water dissociation is suppressed on the polaronic sites.
In the discussions above we have shown that non-polaronic sites on polaronic surfaces can enhance 
water dissociation. 
In the SI we show that such effects remain in hydrogen bonded systems.

Let us now connect our results to experiments.
First, there is already support for one aspect of our observations in the recent experiments.
In a previous study, some of us observed the segregation of polarons towards 
the water covered rutile (110) surface using photoemission spectroscopy \cite{yim_visualization_2018}.
The observations indicate an attractive water-polaron interaction, which is consistent with the enhanced water adsorption and enhanced HB strength shown in this study.
Second, our results might help to rationalize some of the debates associated with water on $\text{TiO}_\text{2}$.
Given the ability of polarons to both inhibit and enhance dissociation and the fact that 
polarons are mobile and that this mobility will depend on temperature it is easy to see how 
different measurement could yield different conclusions in terms of water dissociation \cite{schaub_oxygen_2001, wendt_formation_2006,bikondoa_direct_2006,walle_experimental_2009,duncan_water_2012,diebold_perspective:_2017, wang_probing_2017}.
In addition, the interfacial structure at the rutile (110) water interface is still highly debated 
(see e.g. Refs. \onlinecite{hussain_structure_2017, balajka786}).
Understanding the polaronic effects on the adsorption of water/hydroxide and the strength of HBs may shed light on this issue that the stability of interfacial structure might be affected by polarons induced in experiments. 
We are certainly not saying that polarons are the full story in terms of the water structure and dissociation debates, however our results suggest that they are likely to be more relevant than previously anticipated. 

Besides rutile, which has been the focus of this study, 
there is another important phase of $\text{TiO}_\text{2}$, namely
anatase, which is generally regarded as a more efficient photocatalyst than 
rutile 
\cite{tanaka_effect_1991}.
However, surface science techniques have not shown a clear preference for water dissociation on pristine anatase surfaces compared to pristine rutile surfaces.
The higher catalytic activity of anatase has been linked to factors such as higher electron mobility, longer electron-hole pair lifetimes, and a larger band gap \cite{tang_electrical_1994, xu_photocatalytic_2011}.
This study, complementing Selcuk and Selloni's study of anatase \cite{selcuk_facet-dependent_2016}, suggests polaronic effects should be considered as additional possibilities concerning the difference between rutile and anatase.
In general, there are many other polaronic materials and complex surface phenomena where the interplay and links between polarons and adsorbates have yet to be established.
In the SI we show such effects can be extended to the dissociation of methanol on rutile, 
a system of importance to heterogeneous catalysis \cite{silber_adsorbate-induced_2016}.
On other polaronic substrates, the binding of a water molecule to polaronic sites will likely lead to the suppression effect observed here, but more complex behavior is expected at non-polaronic sites and where the HB network is different.
In addition, polarons may affect the proton dynamics in the contact layer of the water $\text{TiO}_\text{2}$ interface, thus having a strong effect on the reaction dynamics.
The dynamics of polarons may also affect the adsorption, hydrogen bonding and dissociation, which should be further investigated in the future.

To conclude, this study has revealed various key roles played by polarons in the rutile (110) surface on water adsorption, dissociation and hydrogen bonding at the surface. 
Firstly, such polarons significantly modulate the adsorption of water and hydroxide groups on the surface.
Consequently they suppress the dissociation of adsorbed water molecules on polaronic sites, while on non-polaronic sites the effects are less significant, and even a slight enhancement is possible.
The suppressing effect extends to higher water coverage and aqueous interfaces.
In addition, we find water molecules and hydroxide groups become better HB acceptors when bonded to polarons.
Overall, our study should encourage further high-resolution experiments and advanced theoretical simulations to disentangle the key polaronic features affecting heterogeneous physical systems and processes.

\section*{Acknowledgements}
We thank Geoff Thornton for helpful discussions.
J.C., C.P. and A.M. are grateful to the Alexander von Humboldt Foundation for post doctoral
research fellowships and a Bessel Research Award, respectively.
J.C., C.P. and A.M. were also supported by the European Research Council 
under the European Union's Seventh Framework Programme
(FP/2007-2013) / ERC Grant Agreement number 616121 (HeteroIce project).
We thank the High-performance Computing 
Platform of Peking University, the UKCP consortium (EP/F036884/1), the
UK  Materials  and  Molecular  Modelling  Hub (EP/P020194/1), and the Max-Planck-Gesellschaft for computational resources.


%

\end{document}